\documentclass[a4paper,11pt]{article}
\pdfoutput=1
\usepackage{jheppub_kim}
\usepackage[T1]{fontenc}
\usepackage{subfigure}

\usepackage{multirow, graphicx,amssymb,url,mathrsfs,amsmath}
\usepackage{wrapfig,boxedminipage,subfigure,epsfig}
\usepackage{amsxtra,amstext,latexsym,dsfont,amsfonts}
\usepackage{color}
\usepackage[dvipsnames]{xcolor}
\usepackage{float}
\usepackage{slashed}
\usepackage{calligra}
\DeclareFontShape{T1}{calligra}{m}{n}{<->s*[2.2]callig15}{}
\DeclareMathAlphabet{\mathcalligra}{T1}{calligra}{m}{n}

\usepackage{tikz}
\usetikzlibrary{calc,positioning}
\usetikzlibrary{decorations.pathreplacing}

\definecolor{shadecolor}{rgb}{0.9,0.9,0.95}

      \renewcommand{\b}{\beta}

\newcommand{\be}{\begin{equation}}
\newcommand{\ee}{\end{equation}}
\newcommand{\bea}{\begin{eqnarray}}
\newcommand{\eea}{\end{eqnarray}}

\usepackage{comment}

\newcommand{\abs}[1]{\left| #1 \right|}

\title{\boldmath AdS/Deep-Learning made easy: simple examples}

\author[a]{Mugeon Song,}
\emailAdd{gx7179@gmail.com}
\author[a, b]{Maverick S. H. Oh,}
\emailAdd{maverick.sh.oh@gmail.com}
\author[a]{Yongjun Ahn,}
\emailAdd{yongjunahn619@gmail.com}
\author[a]{and Keun-Young Kim}
\emailAdd{fortoe@gist.ac.kr}

\affiliation[a]{Gwangju Institute of Science and Technology (GIST),\\Department of Physics and Photon Science, Gwangju, South Korea}
\affiliation[b]{University of California--Merced,\\Department of Physics, Merced, CA, USA}

\abstract{

Deep learning has been widely and actively used in various research areas. Recently, in the gauge/gravity duality, a new deep learning technique so called the AdS/DL(Deep Learning) has been proposed~\cite{Hashimoto:2018ftp,Hashimoto:2018bnb}. The goal of this paper is to describe the essence of the AdS/DL in the simplest possible setups, for those who want to apply it to
the subject of emergent spacetime as a neural network. For prototypical examples, we choose simple classical mechanics problems. This method is a little different from standard deep learning techniques in the sense that not only do we have the right final answers but also obtain physical understanding of learning parameters.

} 

\begin{document}

\maketitle
\flushbottom

\section{Introduction}

Machine learning or deep learning~\cite{Hashimoto:2020bnb} techniques have become very useful and novel tools in various research areas.
Recently, an interesting machine learning idea  has been proposed by Hashimoto et al. in \cite{Hashimoto:2018ftp,Hashimoto:2018bnb}, where the authors apply the deep learning techniques to the problems in the gauge/gravity duality~\cite{Zaanen:2015oix,  Ammon:2015wua}. They showed that the spacetime metric can be ``deep-learned'' by the boundary conditions of the scalar field, which lives in that space.

The essential deep learning (DL) idea of \cite{Hashimoto:2018ftp,Hashimoto:2018bnb} is to construct the neural network (NN) by using the structure of the differential equation. The discretized version of the differential equation include the information of physical parameters such as a metric. The discretized variable plays a role of different ``layers'' of the NN and the dynamic variables correspond to nodes. Therefore, training the NN means training the physical parameters so, at the end of the day, we can extract the trained physical parameters. This idea is dubbed AdS/DL(Deep Learning).

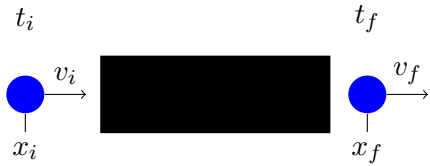
\begin{figure}
    \centering
\begin{tikzpicture}
  \node (a) [blue, fill=blue, circle, minimum size=5mm] at (0,0) {} node at ($(a)+(0,1)$) {$t_i$};
  \draw [black, ->] (a) -- ($(a)+(0.8,0)$) node [midway, above] {$v_i$};
  \draw [black, thick, fill=black] (1,-0.5) rectangle (4,0.5);
  \node (b) [blue, fill=blue, circle, minimum size=5mm] at (4.5,0) {} node at ($(b)+(0,1)$) {$t_f$};
  \draw [black, ->] (b) -- ($(b)+(0.8,0)$) node [midway, above] {$v_f$};
  \draw [black] (a)--($(a)+(0,-0.5)$) node [below]{$x_i$};
  \draw [black] (b)--($(b)+(0,-0.5)$) node [below]{$x_f$};
\end{tikzpicture}
\caption{A ball goes though a ``black-box'' and the velocity of the ball changes from $v_i$ at $t_i$ to $v_f$ at $t_f$. It is very challenging to retrieve the information inside the black-box when the given data is limited by initial and final data.}
\label{fig:blackbox}
\end{figure}

In this paper, we apply the AdS/DL technique to simple classical mechanics problems such as Fig.~\ref{fig:blackbox}. By considering simple examples, 
we highlight the essential idea of AdS/DL without much details of the gauge/gravity duality, which will be useful for those who want to apply this method to the subject of emergent spacetime as a neural network. Furthermore, our work will be a good starting point to learn a physics-friendly NN technique rather than the classical way from computer science.

Let us describe a prototypical problem. Suppose that we want to figure out the force in the black box shown in Fig.~\ref{fig:blackbox}. We are given only initial and final data, for example, the initial and final position \& velocity, $(x_i, v_i)$ and $(x_f, v_f)$. A standard method is to start with an (educated) guess for a functional form of the force (say, $F(x,v)$). One can use this ``trial'' force to simulate the system by solving the Newton's equation. After trial-and-error simulation and comparison with experimental data we may be able to obtain the approximate functional form of the force. However, if the force is complicated enough it will not be easy to make a good guess at first glance, and it will not be easy to modify the trial function in a simple way. In this situations machine learning can be a very powerful method to obtain the force in the black box.

Usually, when there is a big enough input-output data set, classical DL techniques with NN, even without considering the physical meaning of NN or the structure of the problem, can surely make a model that takes input data points and gives matching output values in a trained region, because that is what DL is good at. Having a wide and deep enough feed-forward NN with linear and nonlinear transformations can trivially make such convergence as the Universal Approximation Theorem (UAT) guarantees~\cite{HORNIK1989359,Bishop2006}. Retrieving physical parameters from such a model is not easy because the network in general has little to do with the mathematical structures of models we want to understand.
However, if we build a NN in a way to reflect the mathematical structure of the problem as in the AdS/DL, we are able to retrieve physical information from the model.
In this case, the discretized time ($t$) plays a role of layer and the dynamic variables $(x,v)$ correspond to the nodes.\footnote{For comparison, the variables $(\phi, \pi, \eta)$ in \cite{Hashimoto:2018ftp} correspond to $(x,v,t)$ in section \ref{sec:case1} of this paper. } The unknown force is encoded in the NN so it will be trained.


This paper is orgarnized as follows. In section 2, the general framework of building and training NN from EOM is introduced. In section 3 and 4, example problems are tackled with the methodology described in section 2. Section 3 covers a simpler example with one variable (one-dimensional velocity) while section 4 deals with a problem with two variables (one-dimensional position and velocity).
We conclude in section 5.
\par



%
\par
%


\section{General Framework}
The general framework can be divided into three major parts. First, training data set is generated using the EOM of a system and a widely-used ordinary differential equation solver; we used an adaptive Runge-Kutta method of order 5(4)\footnote{scipy.integrate.ode with the integrator dopri5}~\cite{Hairer1993}. Second, a NN is built from EOM with randomly initialized parameters based on the Euler method. Third, the NN is trained with the training data sets. After these three steps, the resultant learned parameters are compared against the right parameters to see if the learning was successful. The first part, training data set generation, is trivial so that we give an elaboration from the second part.
\par

\par

\subsection{Designing Neural Network from EOM}
\begin{figure}
    \centering
    \includegraphics{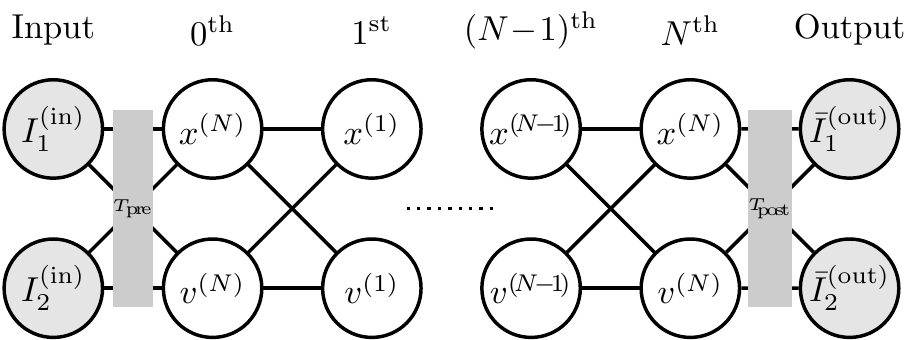}
    \caption{The NN structure with two kinematic variables, $x$ and $v$. Each circular node denotes a neuron with its own variable. The lines between nodes show which nodes are directly correlated with which nodes. Their initial values $(x^{(0)},v^{(0)})$ are calculated from two input information nodes $I_1^{(\text{in})}$ and $I_2^{(\text{in})}$ by a pre-processing transformation $T_{\text{pre}}$. The kinematic variables propagate along the NN from $0^{\text{th}}$ layer to $N^{\text{th}}$ layer with the rule given by the EOM. The final values $(x^{(N)}, v^{(N)})$ are used to calculate the model's output information nodes $\bar{I}_1^{(\text{out})}$ and $\bar{I}_2^{(\text{out})}$, which is compared against the true output values  ${I}_1^{(\text{out})}$ and ${I}_2^{(\text{out})}$ given from the training data set. The number of the nodes in the input and output layers can vary depending on the experimental setup.}
    \label{fig:basicNN}
\end{figure}
In this section, we review how to build a NN from EOM, following the framework suggested by \cite{Hashimoto:2018bnb}. Fig.~\ref{fig:basicNN} shows a basic structure of NN of our interest. It is a feed-forward network, which means the propagation of variables is one-directional without any circular feed-back. Its depth (the number of layers) is set to be $N\!+\!1$ (from $0$ to $N$) excluding the input and output layers, while the width (the number of nodes for a layer) is kept as two. The propagation rule from one layer to the next layer is given by the differential equations from EOM with learning parameters of interest. \par

Here, $T_{\text{pre}}$ is the transformation from the input layer to the $0^{\mathrm{th}}$ layer (pre-processing), which is the identity transformation in our cases, and $T_{\text{post}}$ is the transformation from the $N^{\mathrm{th}}$ layer to the output layer (post-processing). The input layer and the pre-processing transformation is used when there needs a pre-processing of experimental data to the kinematic variables that appear in the EOM. If there is no need for such a pre-processing, one may omit the input layer, which is the case for the rest of this paper. We, however, chose to include the input layer in this section for more general applications which require pre-processing. The output layer corresponds to a set of experimental measurements after the propagation of variables with the EOM. For our cases, it will be final variables and/or flags showing whether or not the trained data points give valid outputs.\footnote{Measurement of velocity using Doppler effect can be a good example of an experimental setup requiring nontrivial pre- and post-processing transformations. In that case, one input/output information node can be initial/final frequency information, while $T_{\text{pre}}$/$T_{\text{post}}$ connects them to initial/final speed values in the NN layers, respectively. }
The details on how we set up the output layer are discussed in the followings sections.
\par

There are two main differences of a NN in our setup from a usual feed-forward NN. First, in our setup, the width of NN stays constant, which is nothing but the number of kinematic variables used in the learning process. In usual cases, however, the width of the NN may vary for different layers to hold more versatility. Second, the propagation rule is set by EOM and there are relatively less learning parameters, whereas most components of propagation rule of a usual NN are set as learning parameters. From NN perspective, our setup may look restrictive, but from the physics perspective, it is more desirable because we may indeed obtain physical understanding on the inner structure of the NN: we want to ``understand'' the system rather than simply having answers. \par


How is the propagation rule given by the EOM? Let us assume that, as time changes from $t_i$ to $t_f$, the following EOM holds
%
\begin{equation}
\ddot{x}=f(x,\dot{x}) \,,
\end{equation}
%
%
or,
\begin{equation} \label{eqn123}
v = \dot{x} \,,  \qquad  \dot{v}=f(x,v) \,.
\end{equation}
If we discretize the time of \eqref{eqn123} and take every time slice as a layer, we may construct a deep neural network with the structure of Fig.~\ref{fig:basicNN} with the following propagation rule, which is essentially the Euler method:
\begin{equation}
\begin{aligned}
x^{(k+1)} &= x^{(k)} + v^{(k)}\,  \Delta t \,,
\\
v^{(k+1)} &= v^{(k)} + f(x^{(k)}, v^{(k)})\,  \Delta t \,,
\end{aligned}
\label{eq:discrete_general}
\end{equation}
where $x^{(k)}$ and $v^{(k)}$ are variables at time $t_i \!+\! k\,\Delta t$ ($k$-th layer)  where $\Delta t := \frac{t_f - t_i}{N}$.\par

Another way of writing \eqref{eq:discrete_general} is separating the linear part and the nonlinear part. The linear transformation can be represented by a weight matrix, $W^{(k)}$ for the $k^{\text{th}}$, while the non-linear transformation is called an activation function, $\varphi^{(k)}$ for the $k^{\text{th}}$, so that the $k^{\text{th}}$ layer variable set,
\begin{equation}
    \mathbf{x}^{(k)}=\left(x^{(k)}, v^{(k)}\right)^T \,,
\end{equation}
propagates to the $(k\!+\!1)^{\text{th}}$ layer by
\begin{equation}
    \mathbf{x}^{(k+1)}=\varphi^{(k)} \left( W^{(k)} \boldsymbol{\mathbf{x}}^{(k)} \right) \,,
\end{equation}
where
\begin{equation} \label{setup123}
    W^{(k)}=\begin{pmatrix}1 && \Delta t \\ 0 && 1
    \end{pmatrix}\,, \qquad  \varphi^{(k)}\begin{pmatrix} a \\b \end{pmatrix} = \begin{pmatrix} a \\ b+f(x^{(k)}, v^{(k)}) \, \Delta t
    \end{pmatrix} \,.
\end{equation}
In this way, the NN is built from EOM and different layers mean different times, except for the input and output layers.
The learning parameters, $f(x^{(k)}, v^{(k)})$ in our case, are randomly set within a reasonable range. The model output ${\mathbf{\bar{x}}}^{(\text{out})}$ can be expressed as follows. To differentiate the true output (training data) from the model output, the model output is specified as variable name with a bar on it whereas the true output without a bar.
\begin{equation}
    \mathbf{\bar{x}}^{(\text{out})} \equiv T_{\text{post}}\bigg(
    \varphi^{(N-1)}
    \Big(W^{(N-1)}
\cdots
    \varphi^{(0)}
    \Big(
    W^{(0)}
    \big(
    T_{\text{pre}} (\mathbf{x}^{(\text{in})})
    \big)
    \Big)
    \Big)
    \bigg)\,,
    \label{eq:outputgen}
\end{equation}
where $\mathbf{x}^{(\text{in})}=\left(I_1^{(\text{in})}, I_2^{(\text{in})}\right)^T$ and  $\mathbf{\bar{x}}^{(\text{out})}=\left(\bar{I}_1^{(\text{out})}, \bar{I}_2^{(\text{out})}\right)^T$. The true output from the training data is denoted as $\mathbf{x}^{(\text{out})}=\left(I_1^{(\text{out})}, I_2^{(\text{out})}\right)^T$.
\par

\subsection{Training Neural Network}
Note that the weight matrix $W^{(k)}$ and the activation function $\varphi^{(k)}$ of the NN is constructed according to the EOM as shown in \eqref{setup123}. Thus,
our goal is to train the function $f(x^{(k)}, v^{(k)})$ using the NN and input/output data.
A single pair $(\mathbf{x}^{(\text{in})}, \mathbf{x}^{(\text{out})})$ is called a training data point, and a whole collection of them $\{(\mathbf{x}^{(\text{in})}, \mathbf{x}^{(\text{out})})\}$ is called a training data set. From the training data set, one can define a error function (a.k.a. loss function) as
\begin{equation}
    E = \frac{1}{n_\text{batch}}\sum_{\text{batch}} \abs{\bar{\mathbf{x}}^{(\text{out})}-\mathbf{x}^{(\text{out})}} + E_{\text{reg}}\,,
    \label{eq:lossfunc}
\end{equation}
where a batch is a part of data set chosen for one learning cycle and $n_{\text{batch}}$ is the number of data points for one batch.
For example, if there are 500 data points in total and 100 data points are used for one batch of the learning process, $n_{\text{batch}}$ is $100$ and five learning cycles cover the whole data set, which is called one epoch of learning.
The summation over ``batch'' means that we add up the term from every data point from the batch.
Dividing the data set into batches makes the learning process more efficient especially when the data set is big.
To make multiple parameters optimized with enough stability, many epochs of learning is required.\par

The first term in \eqref{eq:lossfunc} is the $\text{L}^1$-norm error of the batch calculated from the difference of the output from the NN model $\bar{\mathbf{x}}^{(\text{out})}$ and the true output from the training data set ${\mathbf{x}}^{(\text{out})}$, which is one of the most widely-used error functions. The second term, $E_{\text{reg}}$, is the regularization error which makes unphysical solutions (e.g. unnecessarily zigzagging solutions) unfavorable in learning. The details on $E_{\text{reg}}$ will be delivered in the following sections. Note that the error function defined here is one example of possible choices. The structure of $E$ can vary depending on the nature of problems. Please refer to Sec.~\ref{sec:case1} for a variation.

In general, the value of $E$ depends on both the weight matrix and the activation function.\footnote{In a usual NN, the activation function is fixed as a nonlinear function, such as a sigmoid function or a rectified unit function, and the weight matrix is trained.} For our model, however, the weight matrices are constant in the sense that they are not learning parameters in the NN so that the activation functions $\varphi^{(k)}$, or more specifically the parameters $f(x^{(k)}, v^{(k)})$, are the only parameters to be learned while minimizing the value of $E$. As an optimizer (learning mechanism), the two most classic choices are stochastic gradient descent (SGD) and Adam, where the former is more stable and the latter is faster in many cases~\cite{Kingma2015AdamAM}. We used Adam method with Python 3 and PyTorch as a general machine learning environment.

\par

\section{Case 1: Finding a Force $F_1(v)$}
In this section, we describe the basic idea of our method by using one of the most simplest examples. Here, we use only one kinematic variable, the one-dimensional velocity $v$, to extract information of the velocity-dependent drag force $F_{1,\text{True}}(v)$ of a given system. This example is very simple but the application of DL methodology is relevant and clear. The drag force is designed to be non-trivial to fully test the capability of the methodology.

\paragraph{Problem definition}

\begin{figure}
    \centering
\begin{tikzpicture}
  \draw [gray, thick, fill=gray] (-0.5, 0) rectangle (0.5, -5.5);
  \node (a) [draw=blue, dotted, fill=none, line width=1, circle, minimum size=5mm] at (0,-0.5) {} node at ($(a)+(-1,0)$) {$t_i$};
  \draw [black, ->] (a) -- ($(a)+(0,-0.8)$) node [midway, right] {$v_i$};
  \node (b) [draw=blue, dotted, fill=none, line width = 1, circle, minimum size=5mm] at (0, -4.5) {} node at ($(b)+(-1,0)$) {$t_f$};
  \draw [black, ->] (b) -- ($(b)+(0,-0.8)$) node [midway, right] {$v_f$};
 \draw [thick, black, ->] (1,-2) -- (1,-4) node [midway, right] (g) {$g$};
 \draw [thick, black, ->] (0,-3.5) -- (0,-2.5) node [midway, left] (Fv) {$F_1$};
\end{tikzpicture}
\caption{The problem setup of case 1. A ball in a known constant gravitational acceleration $g$ downward goes though a ``black-box'' filled with homogeneous medium with an unknown drag force $F_1(v)$. From experiments, multiple initial and final velocity values, $v_i$ and $v_f$, are recorded at fixed initial and final time $t_i$ and $t_f$.}
\label{fig:blackbox_2}
\end{figure}
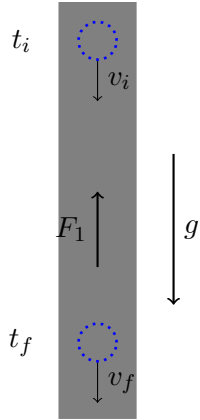
We consider a problem setup described in Fig.~\ref{fig:blackbox_2}. A ball with mass $m$ is dropped with the initial velocity $v_i$ at time $t_i$ through a medium with an unknown complicated drag force $F_1(v)$ under a constant gravitational acceleration $g$ downward.
At time $t_f$, the velocity $v_f$ is recorded. The times $t_i$ and $t_f$ are fixed whereas $v_i$ varies as well as $v_f$ so that we have the input-output data set $\left\{\left(v_i,\, v_f\right)\right\}$ for training. The EOM is given as follows and we want to find the drag force $F_1(v)$:
\begin{equation} \label{eq1}
\dot{v} = - g + \cfrac{F_1(v)}{m} \, .
\end{equation}
\begin{figure}
    \centering
    \includegraphics[width=0.7\textwidth, page={2}]{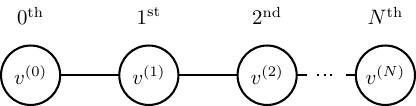}
    \caption{The diagram of the deep neural network for case 1. }
    \label{fig:NN, case1}
\end{figure}

\paragraph{Method}
Because we only have one kinematic variable $v$, it is enough to build a NN with one node per a layer (the width of one) as described in Fig.~\ref{fig:NN, case1}. We omitted the input and output layers in Fig.~\ref{fig:NN, case1}, since the $0^{\text{th}}$ and $N^{\text{th}}$ layer values, $v^{(0)}$ and $v^{(N)}$, are themselves used as the input and output layers, $v_i$ and $\bar{v}_f$, without any pre- or post-processing.
The propagation rule of the NN is written as follows.
\begin{equation} \label{eq2}
v^{(k+1)} = v^{(k)}- \left(  g - \cfrac{ F_1(v^{(k)})}{ m }\right)\,\Delta t \,.
\end{equation}
\par

\begin{figure}
\centering
\subfigure[Learning progress of $\bar{v}_f$ with different epochs]
{\includegraphics[width=0.48\textwidth]{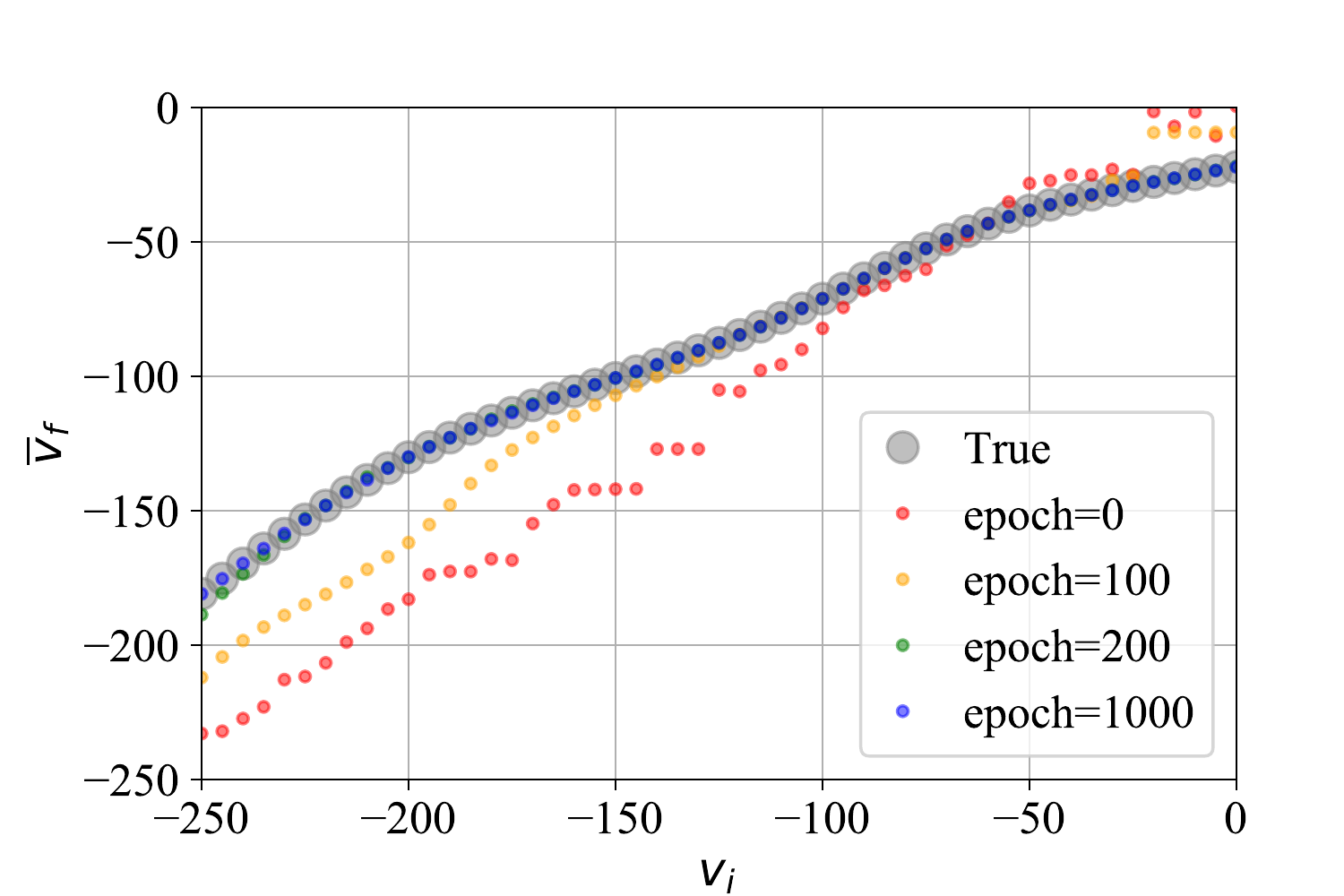}\label{case1a}}
\subfigure[Learning progress of $F_1(v)$ with different epochs]
  {\includegraphics[width=0.48\textwidth]{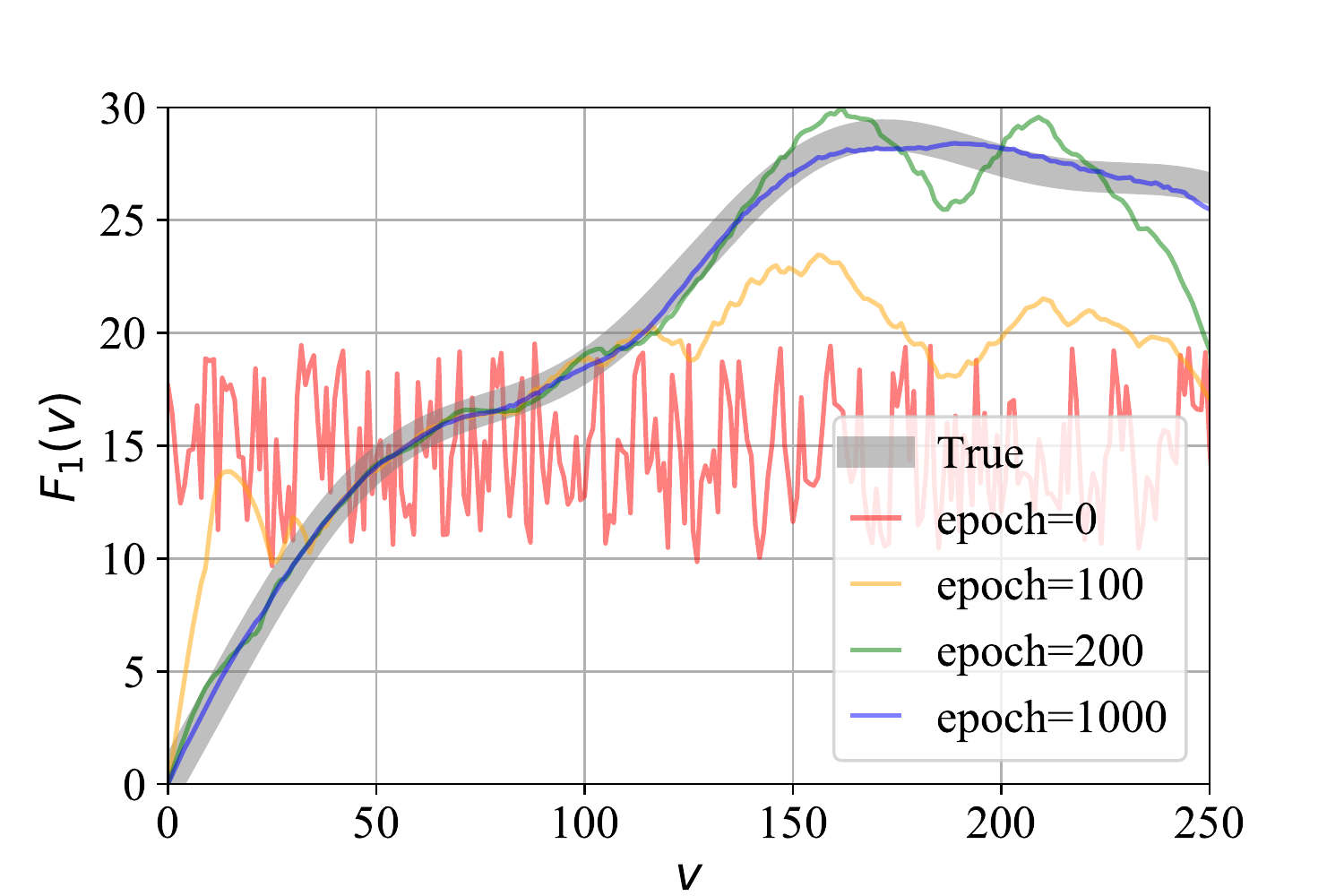}\label{case1b}}
\caption{case 1: comparison of trained data for different epochs and true data. With a big enough epoch, for example 1000, the trained data (blue points/curve) agree with the true data (gray points/curve).}
\label{fig:Case1_trainingdata}
\end{figure}

The initial velocity values are set by $v_i\in[-250, 0]$, evenly spaced with the gap of 5 (i.e. $-250, -245, \cdots, 0$) and the corresponding $v_f$ is calculated from an ordinary differential equation (ODE) solver independent from the NN, which is shown as thick gray points in  Fig.~\ref{case1a}.
Thus, the total number of collected data points ($v_i, v_f$) is $n_{\text{data}}=51$.\par

As mentioned above, it is possible to build a NN with one kinematic variable $v$ and learn $F_1$ from the training data set. The depth of NN is set by $N=10$. The drag force $F_1$ is modeled as an array of size $L=251$, where its $i^{\text{th}}$ element $F_{1, i}$ corresponds to the value of drag force when $\abs{v}=i$ ($i=0, 1, 2, \cdots, 250$); $F_{1,i}=F_1(i)$. The array can hold the information of the drag force for integer speed values $\abs{v}\in [0, 250]$ where $250$ is the upper limit of the speed of ball during the data collection with the true drag force $F_{1, \text{true}}(v)$.\footnote{During the learning process, however, some data points can have $\abs{v}>250$ by chance because of their initial random drag force profile. In that case, they referred to the drag force value $F_1(v=250)$.} When the speed is not an integer, which is true for most cases, the value is linearly interpolated from two nearest integer values. For example, if $v=0.4$, the drag force value is calculated by $F_{1} (0.4)=(1-0.4)\times F_{1}(0) + 0.4 \times F_{1} (1)$. \par

Our goal is to train $F_1$ to yield  $F_{1,\mathrm{True}}$. Let us now refine our notation by adding the superscript $(j)$ to denote the intermediate outputs by $F_1^{(j)}$. The initial drag force $F_1^{(0)}$ is set by $L=251$ uniform random numbers between $(10, 20)$, as a ``first guess''. See the red wiggly line in Fig.~\ref{case1b}.
The $L$ elements of the drag force array are learning parameters, which are updated in the direction of reducing the value of the error function. The error gets minimized as learning proceeds, and the error at $j$-th learning cycle $E^{(j)}$ is:
\begin{equation}
    E^{(j)} = \cfrac{1}{{n_{\text{batch}}}} \sum_{\text{batch}} \abs{\bar{v}^{(j)}_{f} - v_{f}} + \left(F_{1}^{(j)}(0)\right)^2 + c_{1} \sum_{i=0}^{L-1} \left(F_{1}^{(j)}(i+1) - F_{1}^{(j)} (i)\right)^2 \,.
    \label{eq:losscase1}
\end{equation}
Here, the first term is the $\text{L}^1$-norm error to train the parameters to match the model output of final velocity values, $\bar{v}_f=v^{(N)}$, with the true final velocity values, ${v}_f$, for a given batch input $v_i$. Meanwhile, the number of data points is small enough in this case, so we choose to use the whole data set for every learning cycle; $n_{\text{batch}}\!=\!n_{\text{data}}\!=\!51$.
To put a preference on a physically sensible profile of $F_1$, two regularization terms are introduced. The first term, $\left(F_{1}(0)\right)^2$ reflects a physical requirement: $F_{1}(0)=0$ which means there should be no drag force when $v=0$.  The second term, $c_{1} \sum_{i=0}^{N-1} \left(F_{1, i+1} - F_{1, i}\right)^2$, is a mean squared error between adjacent $F_1$ array values which gives a preference on smoother profiles; it is not plausible for the drag force to have a spiky zigzag profile. In our computation $c_{1}=0.03$ is used and we explain how to choose a proper value of $c_{1}$ at the end of this section.
As an optimizer, Adam method is used.\footnote{With the learning rate of $0.4$.} For numerical work, we chose $m=1$, $t_i=0$, $t_f=4$, $g=10$.

\paragraph{Examples}
As an example force, the following hypothetical (complicated) form of $F_{1,\text{True}}$ is assumed and the training data set $\left\{\left(v_i, v_f\right)\right\}$ is collected.
\begin{equation} \label{eq111}
    F_{1,\mathrm{True}}(v)=\frac{v (300-v) }{1000}  \left[1+\frac{1}{10} \sin\left(\frac{v}{20}\right)+\frac{1}{10} \cos\left(\frac{v}{40}\right)\right] + \left(\frac{v}{70}\right)^2 \,,
\end{equation}
which is shown as the gray line in Fig.~\ref{case1b}.

The learning result is shown in Fig.~\ref{fig:Case1_trainingdata}. In Fig.~\ref{case1a}, the model output set $\left\{\bar{v}_f\right\}$ is shown with the training data set $\left\{v_f\right\}$ with different epoch numbers. The NN model learns how to match those two precisely by modifying the learning parameters $F_1(v)$ as the number of epochs increases. How $F_1(v)$ is trained over different epoch numbers is shown in Fig.~\ref{case1b}. As these plots show, the NN model matched $\left\{\bar{v}_f\right\}$ with $\left\{v_f\right\}$ accurately and discovered $F_{1,\text{True}}$ profile with high accuracy with a big enough epoch number.
\par
\begin{figure}
    \centering
    \includegraphics[width=0.48\textwidth]{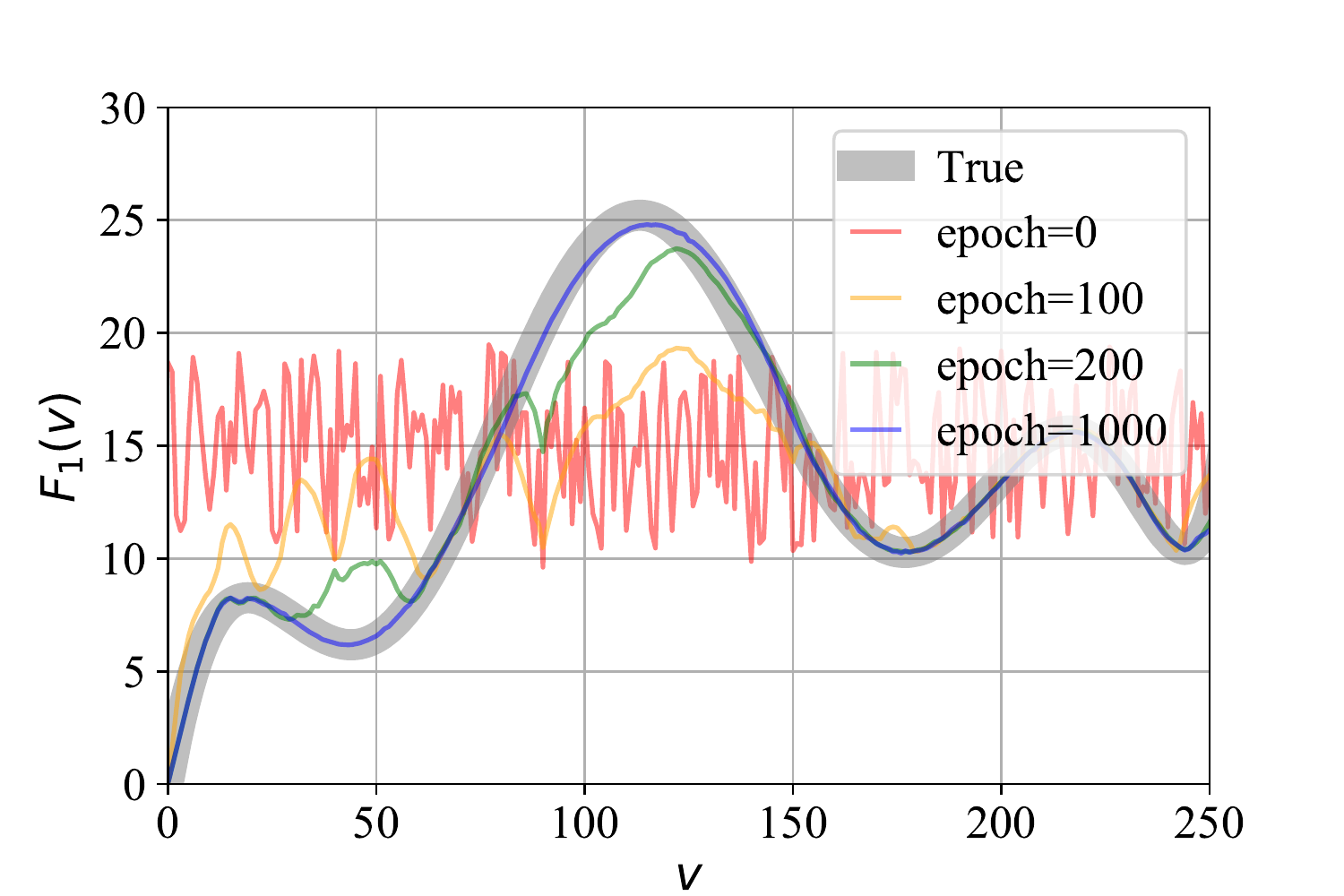}
    \includegraphics[width=0.48\textwidth]{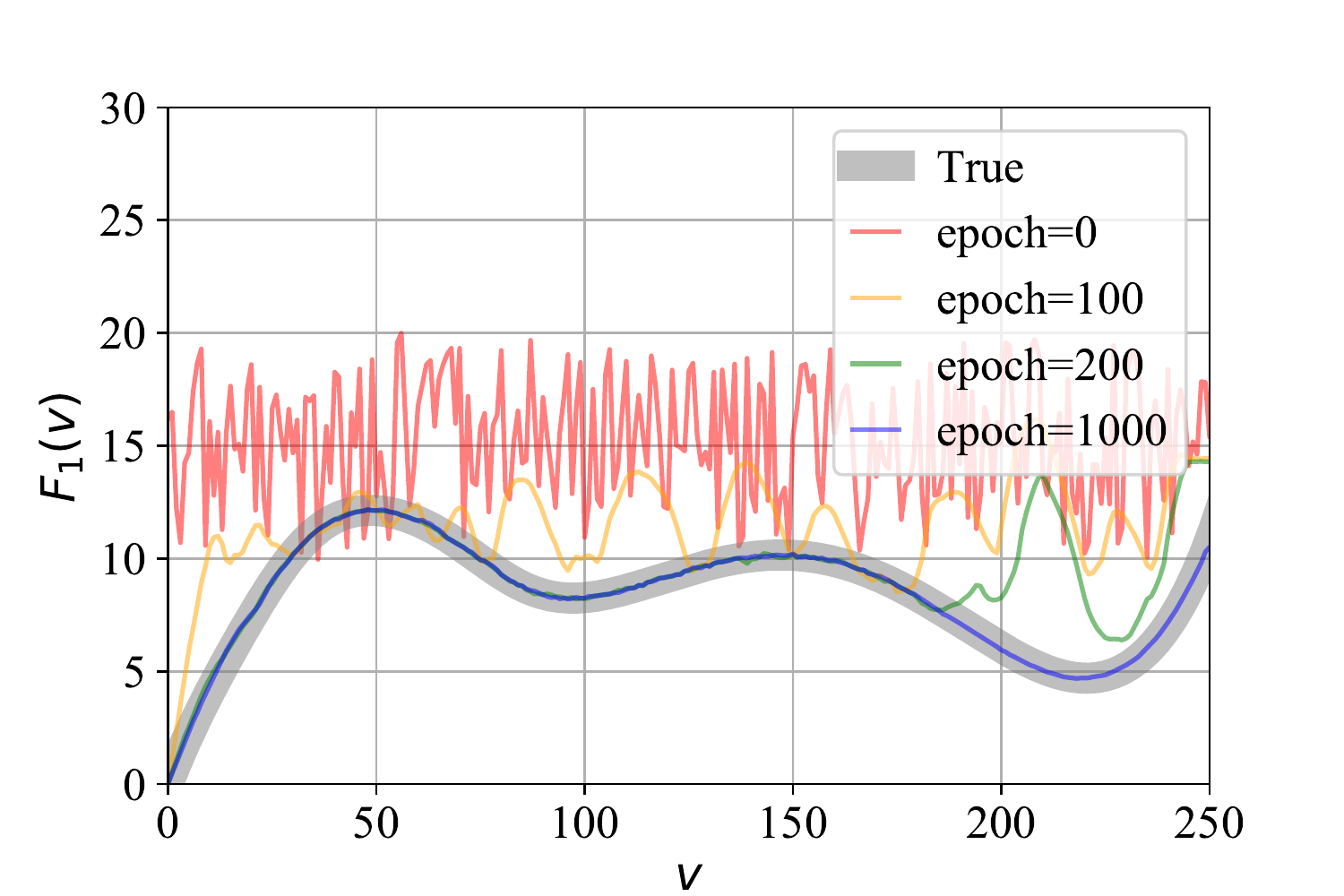}
    \caption[Caption for LOF]{Learning results from different $F_{1,\text{True}}$ profiles.}  \label{case1others}
\end{figure}

To further test the capability of the NN to discover the drag force, different-shaped drag force profiles are tested with the same scheme. As Fig.~\ref{case1others} shows, the NN discovered the right $F_{1,\text{True}}$ profiles accurately as well at $\text{epoch}=1000$.
\footnote{For the purpose of the test of our method, these force profiles are generated by fitting artificially chosen complicated data. Their functional are $1.05135v - 4.24475\times10^{-2}v^2  +5.03648\times10^{-4}v^3 +2.73048 \times 10^{-6}v^4 - 9.34265 \times 10^{-8}v^5  + 6.91675 \times 10^{-10} v^6  -2.16279 \times 10^{-12} v^7 + 2.50634 \times 10^{-15} v^8$, and
$(2.898644\times10^{-1}v -8.043560\times10^{-3}v^2 + 9.985840\times10^{-5}v^3
-5.537040\times10^{-7}v^4 + 1.284692\times10^{-9}v^5
- 8.786800\times10^{-13} v^6)\left(\tanh(\frac{v-25}{20}) + \tanh(\frac{75-v}{20})+1.5\right) $ respectively.}
From the figures, it is clear that both regularization terms (one for setting $F_v(0)=0$ and the other for smoothness) are guiding the learning correctly by filtering out unphysical solutions.

\begin{figure}
\centering
\subfigure[Trained drag force profiles with different $c_{1}$'s.]
{\includegraphics[width=0.48\textwidth]{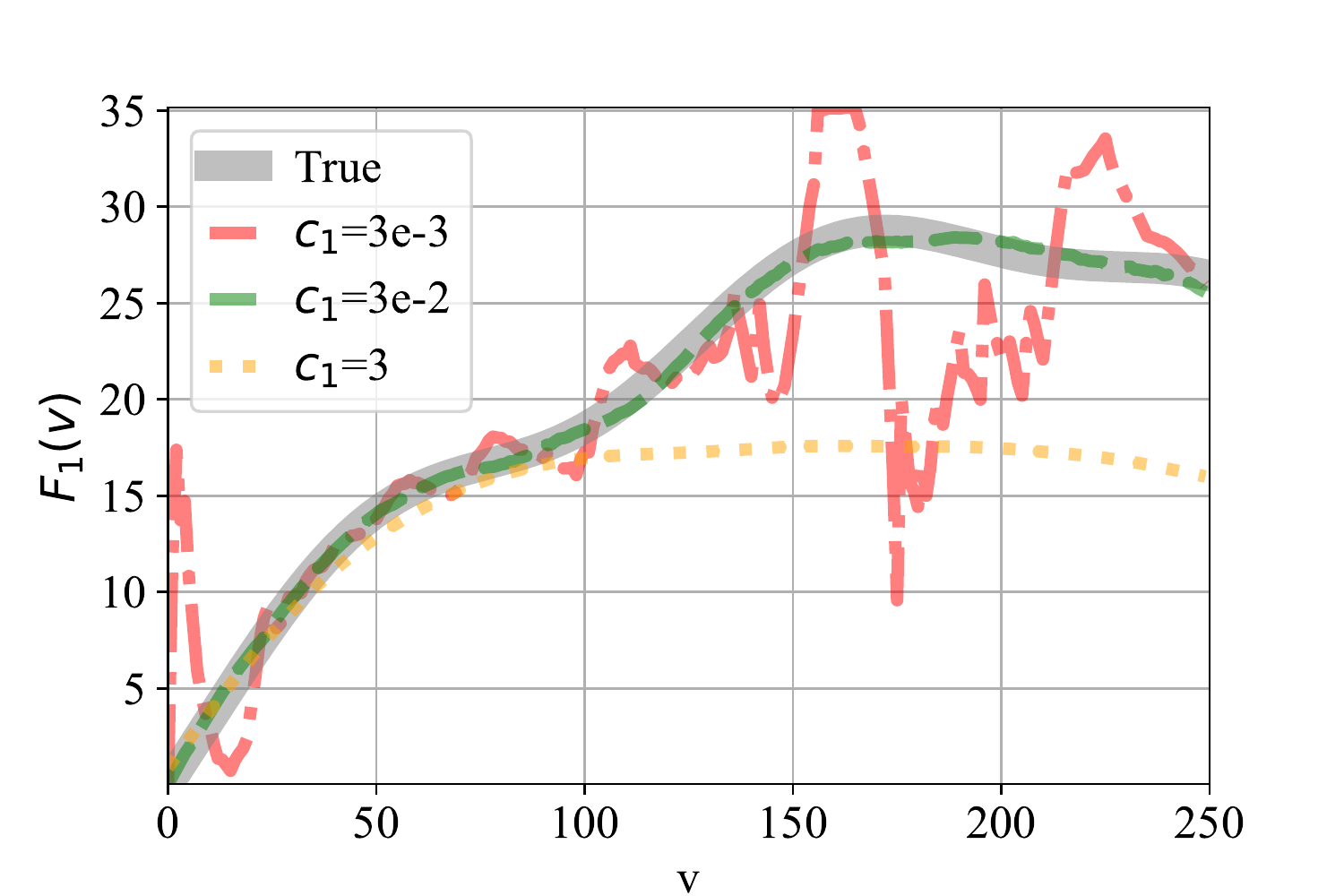} \label{c1a}}
\subfigure[Decreasing tendency of $E$ for different $c_{1}$'s.  ]
{\includegraphics[width=0.48\textwidth]{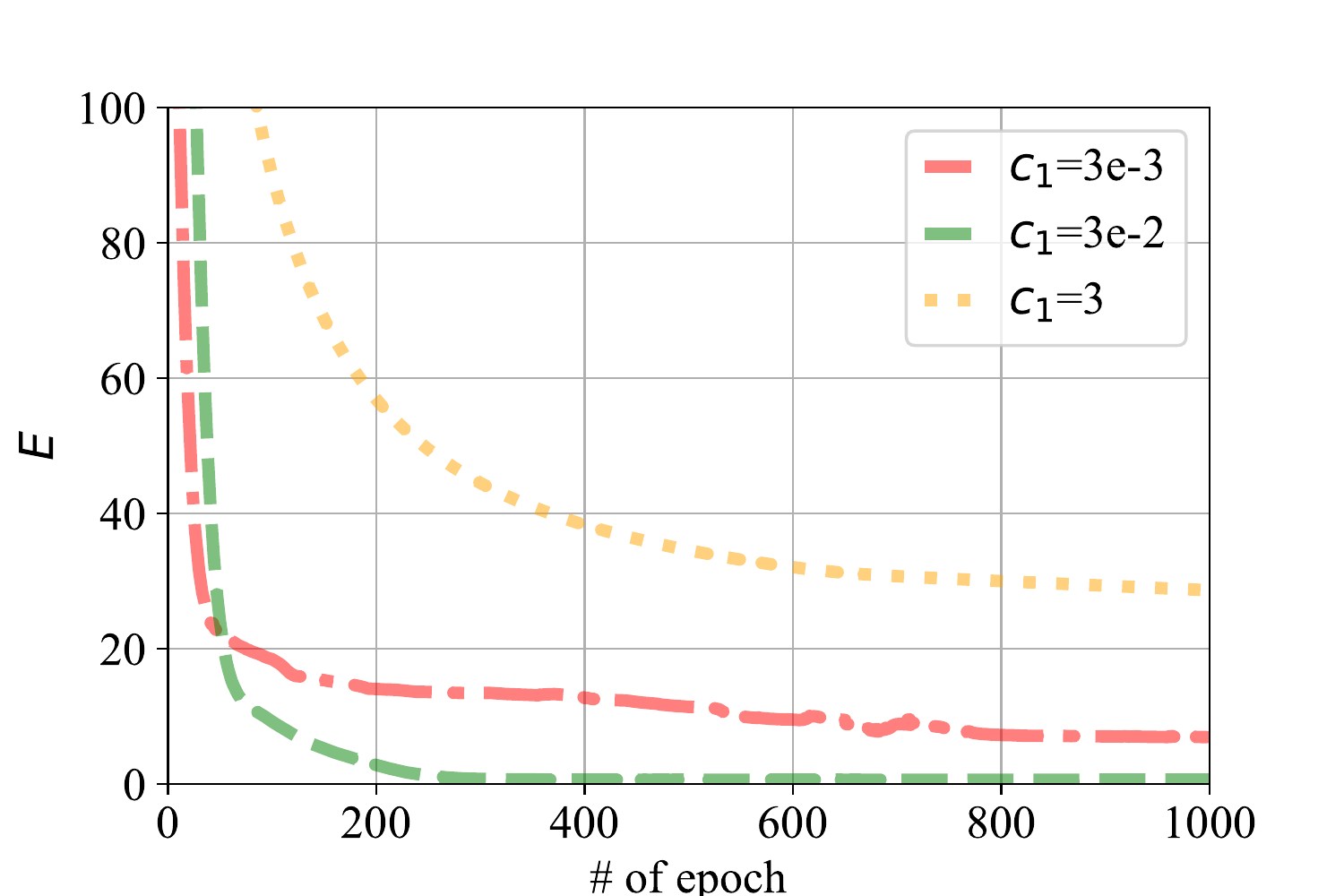}\label{c1b}}
\caption{The regularization $c_1$ dependence of the training results. (a) If $c_{1}$ is too small the force profile is not smooth enough and if $c_{1}$ is too big, the profile becomes flat and deviates from the true profile. (b) We may choose $c_1$ such that the error saturates to a smallest value.}
\label{fig:Case1PenaltyTest}
\end{figure}
We wrap up this section by discussing the choice of $c_1$ for regularization. The value of $c_{1}$ controls the smoothness of the $F_1$ profile. Fig.~\ref{c1a} and \ref{c1b} show the effect of different $c_{1}$ values on the drag force and the error.
If it is too big ($c_1= 3 $, yellow dots), the $F_1$ profile after the learning process ends up being too flat and the error remains very high. If it is too small ($c_1= 0.003$, red dots), the $F_1$ profile stays spiky and the error remains relatively high as well. It turns out that $c_1=0.03$ (green dots) is suitable, showing the trained $F_1$ overlaps very well with $F_{1, \text{True}}$ while resulting in the minimum error. Indeed, this value of $c_1$ can be found by investigating how the error decreases as learning proceeds. As shown in Fig.~\ref{c1b}, the $c_1$ value which gives the minimum error can serve as the best value for regularization.\footnote{This is true in the coarse grain sense. If we want to fine tune the value of $c_1$, we need to be more careful to remove the artificial effect of the regularization term on the entire error.}

\section{Case 2: Finding a Force $F_2(x)$}
\label{sec:case1}
In the second case, two one-dimensional kinematic variables $x$ and $v$ come into play to retrieve the position-dependent force $F_{2, \text{True}}(x)$ of a system from given data. Again, the force is designed to be non-trivial to fully examine the capability of the methodology. The content is divided into three subsections as well; problem definition, method, and examples.

\paragraph{Problem definition}
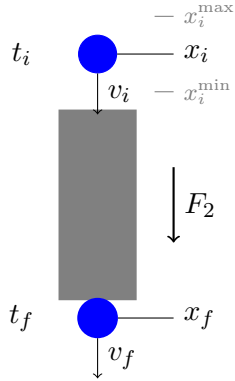
\begin{figure}
    \centering
\begin{tikzpicture}
  \draw [gray, thick, fill=gray] (-0.5, -1.75) rectangle (0.5, -4.25);
  \node (a_down) [draw=none, fill=none, circle, minimum size=5mm] at (0,-1.5) {};
  \draw [gray] ($(a_down)+(0.75,0)$) -- ($(a_down)+(1,0)$) node [right] {\footnotesize\color{gray}{$x_i^{\text{min}}$}};
  \node (a_up) [draw=none, fill=none, circle, minimum size=5mm] at (0,-0.5) {};
  \draw [gray] ($(a_up)+(0.75,0)$) -- ($(a_up)+(1,0)$) node [right] {\footnotesize\color{gray}{$x_i^{ \text{max} }$}};
  \node (a_mid) [draw=blue, fill=blue, circle, minimum size=5mm] at (0,-1) {};
  \draw (a_mid) -- ($(a_mid)+(1,0)$) node [right] {$x_i$};
  \draw [black, ->] (a_mid) -- ($(a_mid)+(0,-0.8)$) node [midway, right] {$v_i$};
   \node at ($(a_mid)+(-1, 0)$) {$t_i$};
  %
  \node (b) [draw=blue, fill=blue, line width = 1, circle, minimum size=5mm] at (0, -4.5) {} node at ($(b)+(-1,0)$) {$t_f$};
  \draw [black, ->] (b) -- ($(b)+(0,-0.8)$) node [midway, right] {$v_f$};
  \draw (b) -- ($(b)+(1,0)$) node [right] {$x_f$};
 \draw [thick, black, ->] (1,-2.5) -- (1,-3.5) node [midway, right] (Fx) {$F_2$};
\end{tikzpicture}
\caption{The problem setup of case 2. A ball goes though a black-box with an unknown force field $F_2(x)$  without any friction. The ball is dropped with speed $v_i$ at time $t_i$ and position $x_i\in (x_i^{\text{min}}, x_i^{\text{max}})$. At $t_f$, if the ball is at the vicinity of $x_f$, a speedometer reads its velocity $v_f$ and the initial kinetic variable set $(x_i, v_i)$ is taken as a positive data point (kind $\kappa\!=\!0$); else, the data point is negative ($\kappa\!=\!1$).}
\label{fig:blackbox_3}
\end{figure}
As shown in Fig.~\ref{fig:blackbox_3}, a ball is shot at the position $x_i$ with the initial velocity $v_i$ at time $t_i$.
The initial position $x_i$ belongs to the range $(x_i^{\text{min}}, x_i^{\text{max}})$ and the initial velocity will be also chosen in certain range so that we can have a window of training data set.

At a fixed final time $t_f$, if the ball is at the vicinity of $x_f$ (\textemdash within $x_f \pm \epsilon$), the initial kinematic variable set $(x_i, v_i)$ is taken as a positive data point (kind $\kappa\!=\!0$) and its velocity $v_f$ is recorded. If the ball is not within $x_f \pm \epsilon$ when $t=t_f$, the data point $(x_i, v_i)$ is taken as a negative one (kind $\kappa\!=\!1$) and we assume that we can not measure a corresponding $v_f$ value. In other words, a positive data point holds two output values $\kappa\!=\!0$ and $v_f$, while a negative point holds one output value $\kappa\!=\!1$.  The EOM is given as $\ddot{x}\!=\!F_2(x)/m$ and we want to find the  force $F_2$. The EOM can be separated into two first order differential equations as follows.
\begin{equation} \label{eq:case2_separate}
\dot{x} = v\,,\qquad
\dot{v} = \frac{1}{m}F_2(x)\,.
\end{equation}
\par
\begin{figure}
    \centering
    \includegraphics[width=0.7\textwidth, page={3}]{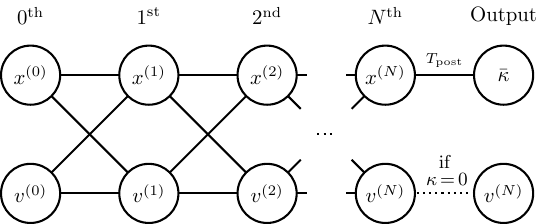}
    \caption{The diagram of the deep neural network for case 2. An input data point at the $0^{\text{th}}$ layer $(x^{(0)}, v^{(0)})$ propagates to the $N^{\text{th}}$ layer. Every data point gives the first output $\bar{\kappa}=T_{\text{post}}\left( x^{(N)} \right)$ while the second output $v^{(N)}$ is given only from one of the positive data points ($\kappa=0$).
    }
    \label{fig:NN, case2}
\end{figure}

\paragraph{Method}
Because we have two kinematic variables $x$ and $v$, two nodes per a layer (the width of two) setup is used for building the NN, as shown in Fig.~\ref{fig:NN, case2}. The input layer is omitted while the output layer is formed using a post-processing transformation $T_{\text{post}}$ on $\bar{x}_f=x^{(N)}$, which judges whether or not $\bar{x}_f$ is at the vicinity of $x_f$ at $t_f$ by $\bar{\kappa}=T_{\text{post}}(\bar{x}_f)\simeq0$ for $\abs{\bar{x}_f-x_f}\le\epsilon$ (model-positive) and $\simeq1$ for $\abs{\bar{x}_f-x_f}\ge\epsilon$ (model-negative). More discussions about $T_{\text{post}}$ follow shortly. For positive data points ($\kappa\!=\!0$), their $\bar{v}_f=v^{(N)}$ values are recorded as well. The propagation rule from EOM is written as follows.
\begin{equation} \label{eq5}
\begin{split}
&x^{(k+1)} = x^{(k)} + v^{(k)} \Delta t \,, \\
&v^{(k+1)} = v^{(k)} + \frac{F_2( x^{(k)} )}{m} \Delta t\, .
\end{split}
\end{equation}
\par

The depth of the NN is set by $N=20$. As in the drag force field $F_1$ of case 1, the force field $F_2$ is modeled as an array that holds the force value for integer positions; $F_2$ is modeled as an array of size $L=21$ covering integer position $x \in [0, 20]$. The $i^{\text{th}}$ component of $F_2$ is the force at $x=i$; $F_{2, i}=F_2(i)$. When a position value is not an integer, the force is linearly interpolated from those of two nearest integer positions.
\par

\begin{figure}[H]
\centering
    \subfigure[Before learning: $\kappa$ and $\bar{\kappa}$ comparison. \newline Blue: positive ($\kappa\!=\!0$), Orange: model-positive\newline ($\bar{\kappa}\!\simeq\!0$), Green: intersection ($\kappa\!=\!0$ and $\bar{\kappa}\!\simeq\!0$).]
    		{\includegraphics[width=0.48\textwidth]{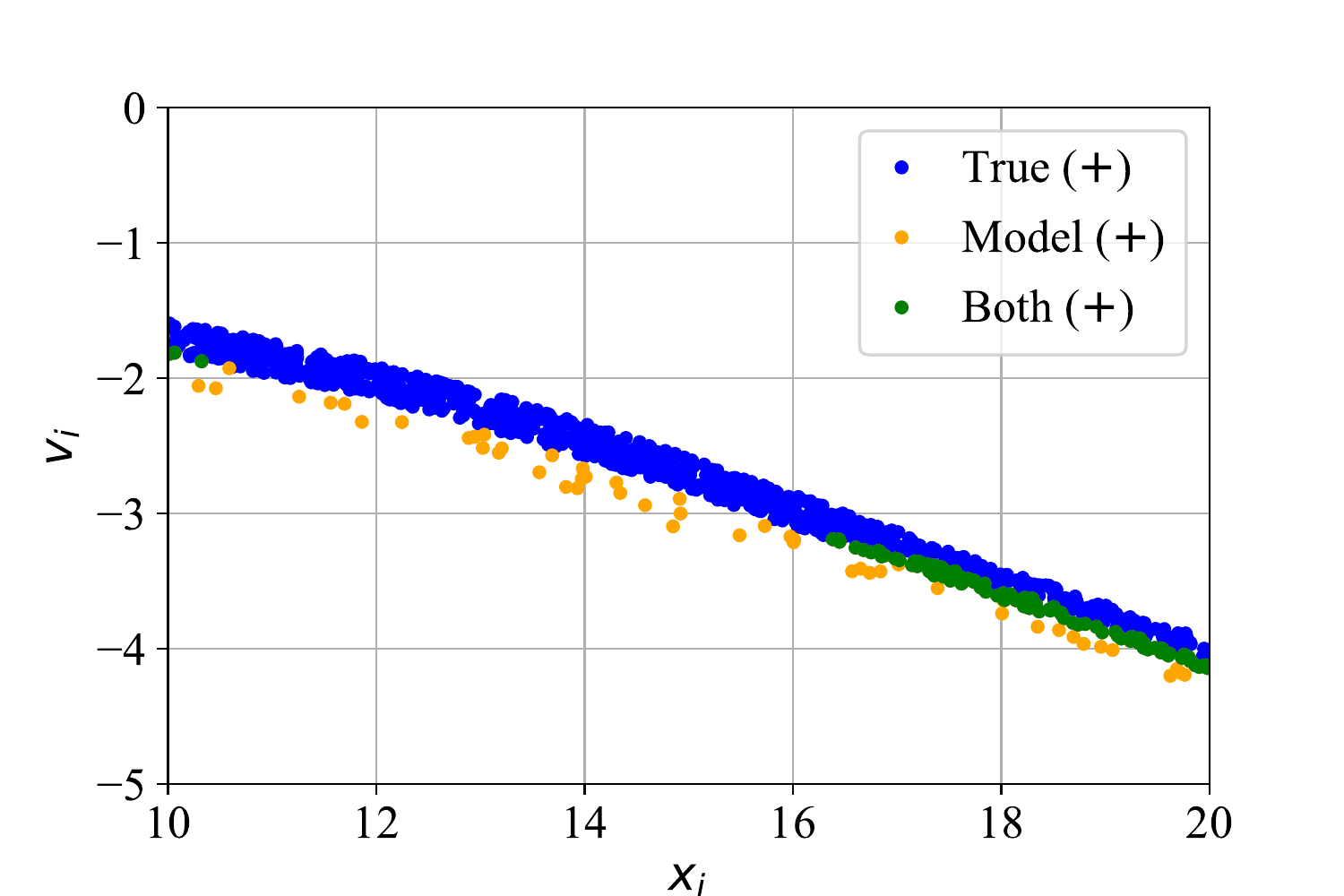}\label{case2traina}}
            \hfill
    \subfigure[Before learning: $v_f$ and $\bar{v}_f$ comparison. \newline Blue: training data ($v_f$), Green: model propagation result ($\bar{v}_f$).]
    		{\includegraphics[width=0.48\textwidth]{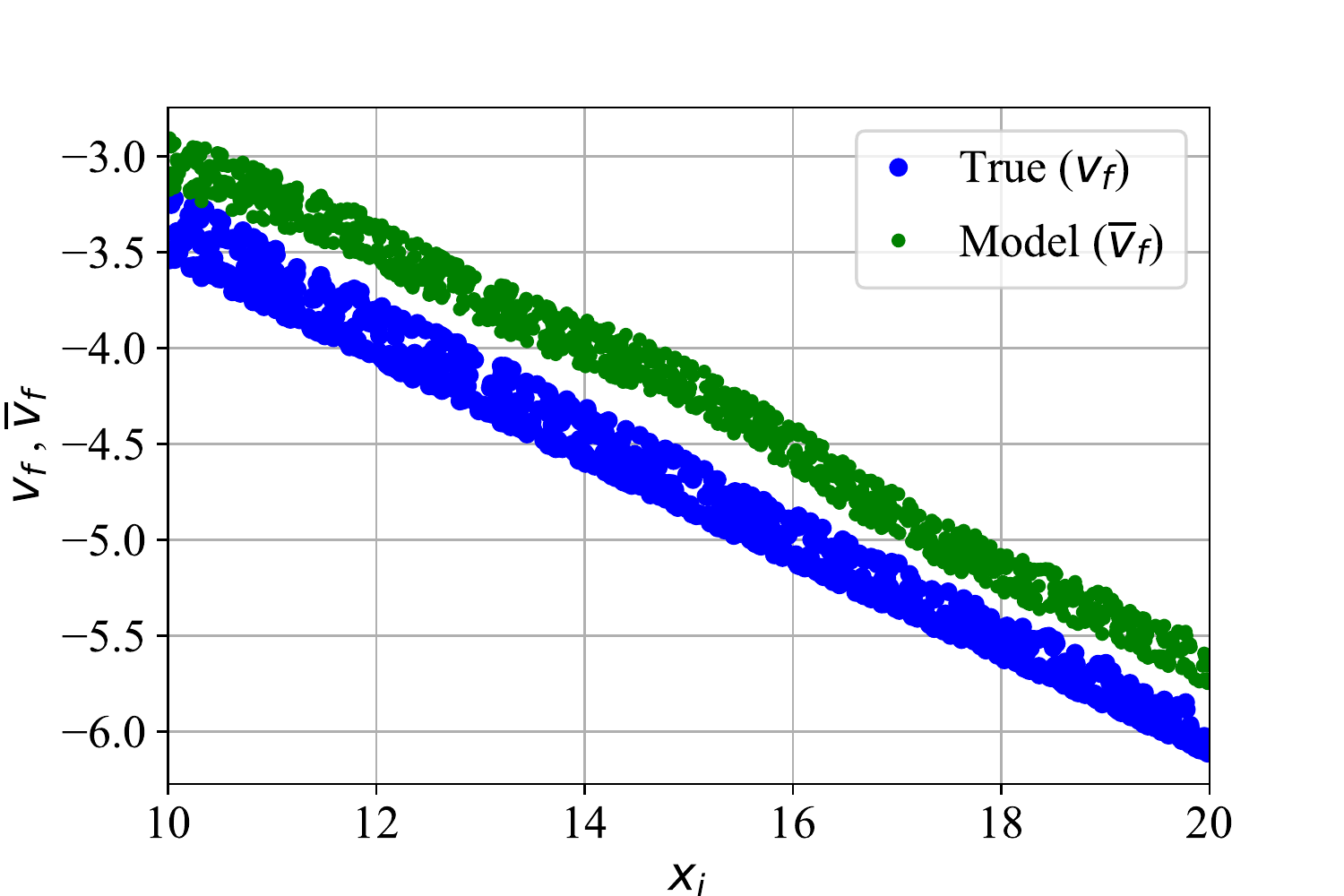}\label{case2trainb}}
            \\
    \subfigure[After learning: $\kappa$ and $\bar{\kappa}$ comparison. \newline Blue: positive ($\kappa\!=\!0$), Orange: model-positive\newline ($\bar{\kappa}\!\simeq\!0$), Green: intersection ($\kappa\!=\!0$ and $\bar{\kappa}\!\simeq\!0$).]
    		{\includegraphics[width=0.48\textwidth]{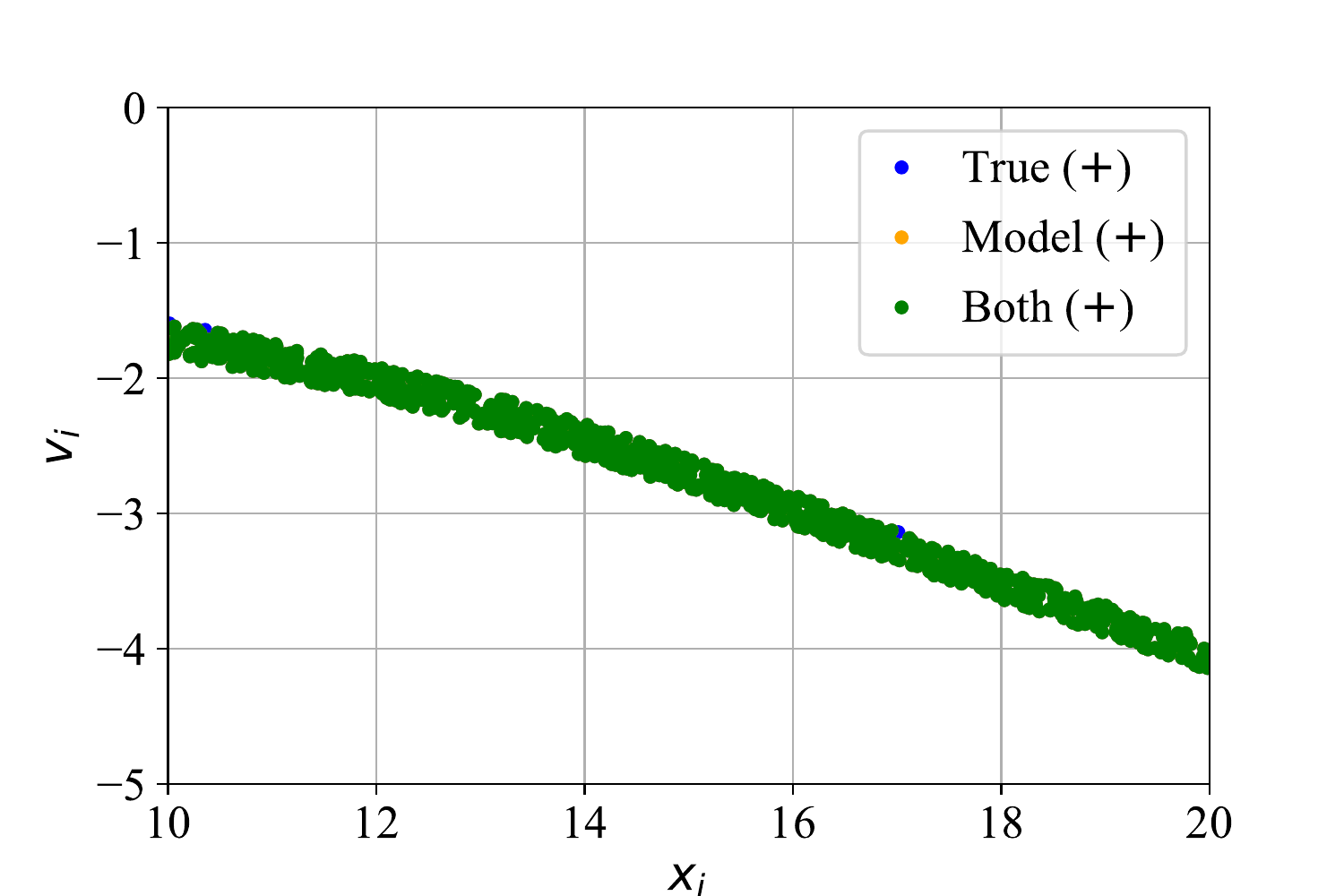}\label{case2trainc}}
            \hfill
    \subfigure[After learning: $v_f$ and $\bar{v}_f$ comparison. \newline Blue: training data ($v_f$), Green: model propagation result ($\bar{v}_f$).]
    		{\includegraphics[width=0.48\textwidth]{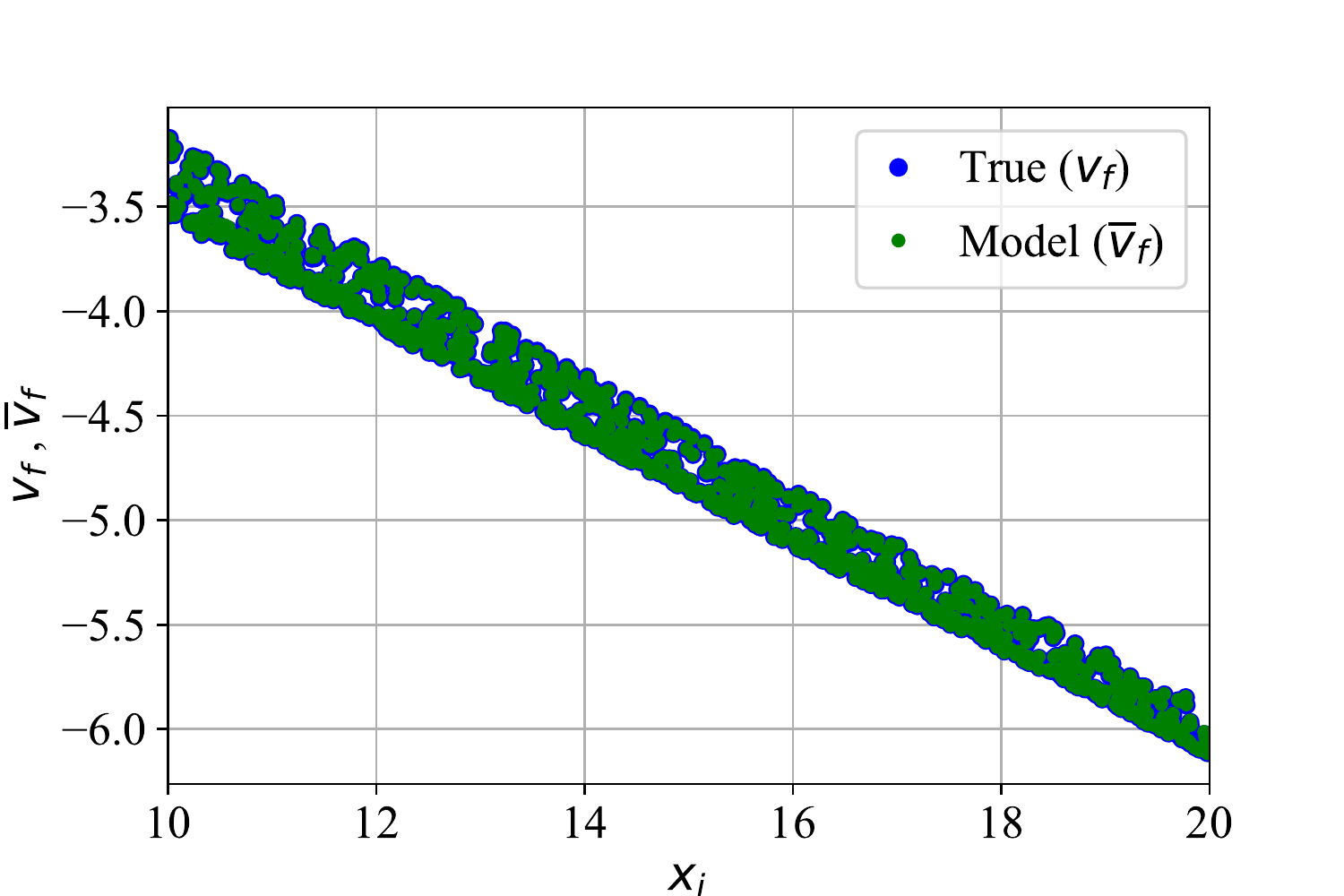}\label{case2traind}}
    \subfigure[Learning progress of $F_2(x)$ with different epochs. \newline Gray: $F_{2,\text{True}}(x)$]{\includegraphics[width=0.48\textwidth]
    {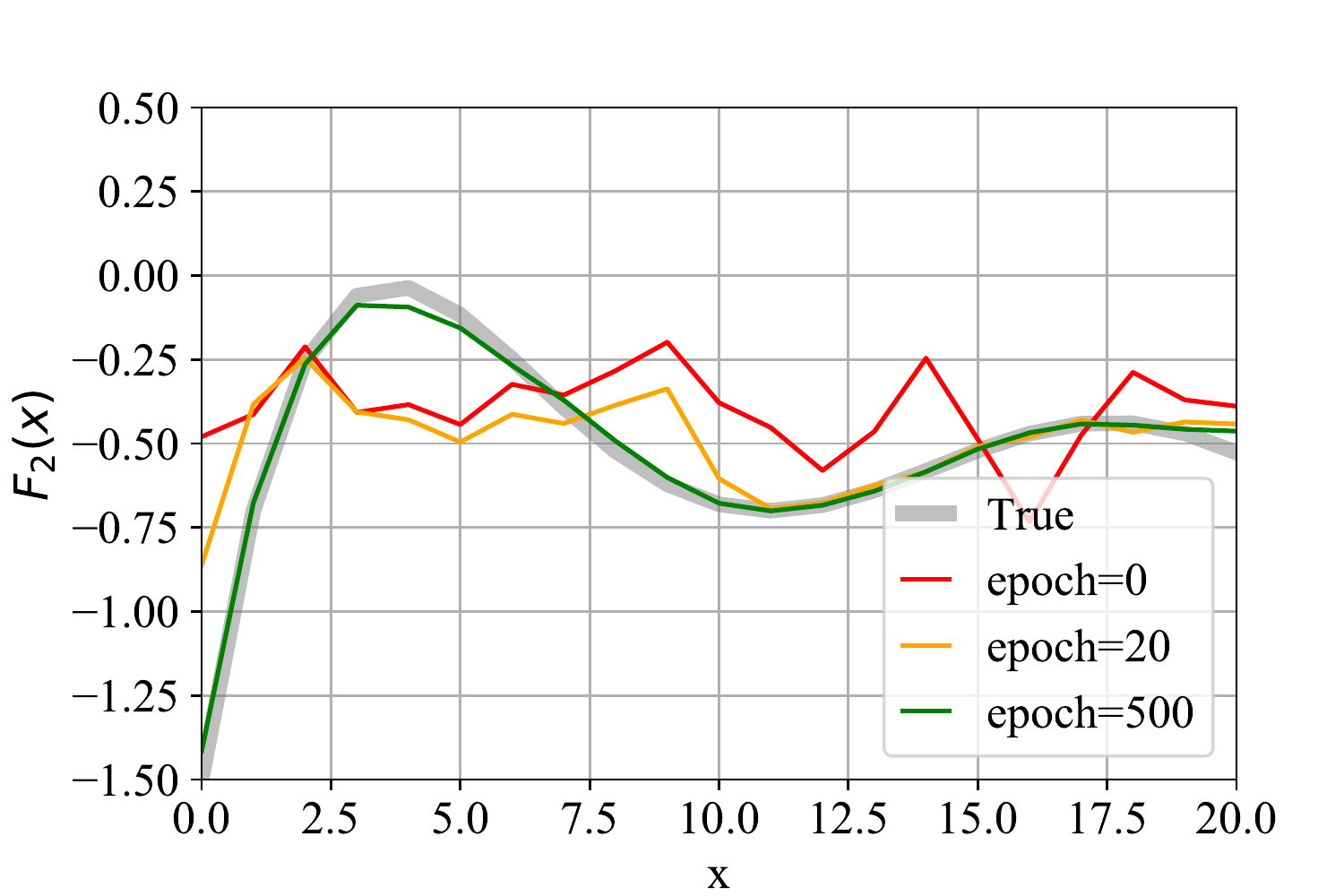}\label{case2traine}}
\caption{Model's output before (a, b) and after (c, d) learning. The learning progress of $F_2(x)$ is shown in (e).
(a) $(x_{i}, v_{i})$ plot of positive and model-positive data points before learning. Most points' $\kappa$ values are incorrectly guessed as $\bar{\kappa}$ by the model NN. (b) $(x_{i}, v_{f})$ and $(x_i,\bar{v}_f)$ plots for positive data points before learning. Most $v_f$ values are incorrectly guessed as $\bar{v}_f$ by the model NN. \newline
(c)  $(x_{i}, v_{i})$ plot of positive and model-positive data points after learning. Most points' $\kappa$ values are correctly learned as $\bar{\kappa}$ by the model NN. (d) $(x_{i}, v_{f})$ and $(x_i,\bar{v}_f)$ plots for positive data points before learning. Most $v_f$ values are correctly learned as $\bar{v}_f$ by the model NN. \newline
(e) As epoch increases, $F_2(x)$ profile approaches to $F_{2,\text{True}}(x)$.}
\label{fig:F_2_train}
\end{figure}

Our goal is to train $F_2$ to yield $F_{2,\text{True}}$. The model's force profile at $j^{\text{th}}$ learning cycle, $F_2^{(j)}$, approaches to $F_{2,\text{True}}$ as $j$ increases, if the learning is correctly designed and performed. The initial $F_2$ array, $F_2^{(0)}$, is set by normal random numbers with the average of $-0.4$ and the standard deviation of $0.1$ as a ``first guess''. See the red wiggly line in Fig.~\ref{case2traine}.
\par

There are two things for the NN model to learn. First, the model needs to distinguish positive and negative data points\textemdash it should match the $\bar{\kappa}$ of a given data point $(x_i, v_i)$ with its actual $\kappa$ properly. Second, the model should be able to match the model's final velocity $\bar{v}_f=v^{(N)}$ with the true final velocity $v_f$ at positive data points $(x_i, v_i, \kappa=0)$.

To reflect these, we need two terms for the error function, one for $\bar{\kappa}$ and $\kappa$ and the other for $\bar{v}_f$ and $v_f$, in addition to the regularization error which gives a preference on smoother profiles. The error function at $j^{\text{th}}$ learning cycle is as follows.
\begin{equation}
\begin{split}
E^{(j)} = \,
   & \mathcal{N}_1 \frac{1}{{n_{\text{batch}}}}
   \sum_{\text{batch}} \abs{\bar{\kappa}^{(j)} - \kappa} +
   \mathcal{N}_2 \frac{1}{n_{\text{batch}, \kappa=0}}
   \sum_{\text{batch}, \kappa=0} \abs{\bar{v}_f^{(j)} - v_f}\\
   &+c_{2} \sum_{i=0}^{L-1} (F_{2, i+1}^{(j)} - F_{2, i}^{(j)})^2  \,,
\end{split}
\label{eq:losscase2}
\end{equation}
where the first and second terms are $\text{L}^1$-norm errors normalized with their size $n_{\text{batch}}$ and $n_{\text{batch}, \kappa=0}$, respectively, scaled by the coefficients $\mathcal{N}_1$ and $\mathcal{N}_2$ that can control relative importance between the two terms (both are set as one for our case). The third term is the regularization error for smoothness of the profile (the mean squared error between adjacent $F_2$ array elements) with the coefficient $c_2$. The model output of the kind variable $\bar{\kappa}=T_{\text{post}}(\bar{x}_f=x^{(N)})$ is calculated using the following post-processing transformation $T_{\text{post}}$, which gives $T_{\text{post}}(\abs{x-x_f}\le\epsilon)\simeq 0$ and $T_{\text{post}}(\abs{x-x_f}>\epsilon)\simeq 1$.
\begin{equation}
\begin{aligned} \label{eq:output}
T_{\text{post}}(\bar{x}_f) &= \cfrac{1}{2} \left(
\tanh{
[20\left(\left(\bar{x}_f-x_f\right) - \epsilon\right)]
}
-
\tanh{
[20\left(\left(\bar{x}_f-x_f\right) + \epsilon\right)]
}
\right)
+ 1\\
&= \cfrac{1}{2} \left(
\tanh{
[20\left(\bar{x}_f - \epsilon\right)]
}
-
\tanh{
[20\left(\bar{x}_f + \epsilon\right)]
}
\right)
+ 1 \qquad {(\because x_f=0)}
\end{aligned}
\end{equation}
The reason to use an analytic function form for $T_{\text{post}}$ rather than a step function is to enable the optimizer to differentiate the error function in the parameter space and find the direction to update parameters to minimize the error; if it is a step function, an optimizer would not be able to find the direction to update the parameters. For learning setup, following values are used: $c_2=0.003$, $n_{\text{batch}}=200$,  $n_{\text{batch}, \kappa=0}=100$, $\epsilon=0.5$. The termination condition was $\text{epochs}= 500$. As an optimizer, Adam method is used.\footnote{With the learning rate of $1\times10^{-2}$.} For numerical work, we chose $m=1$, $t_i=0$, $t_f=4$, $(x_i^{\text{min}}, x_i^{\text{max}})=(10, 20)$, $x_f=0$, and $v_i \in (-5,0)$, which is an enough range of velocity to collect positive data points in our setting.
\par

\begin{figure}[]
\centering
    \subfigure[Blue dots are positive data ($\kappa=0$) and orange dots are negative data ($\kappa=1$). ]{\includegraphics[width=0.48\textwidth]{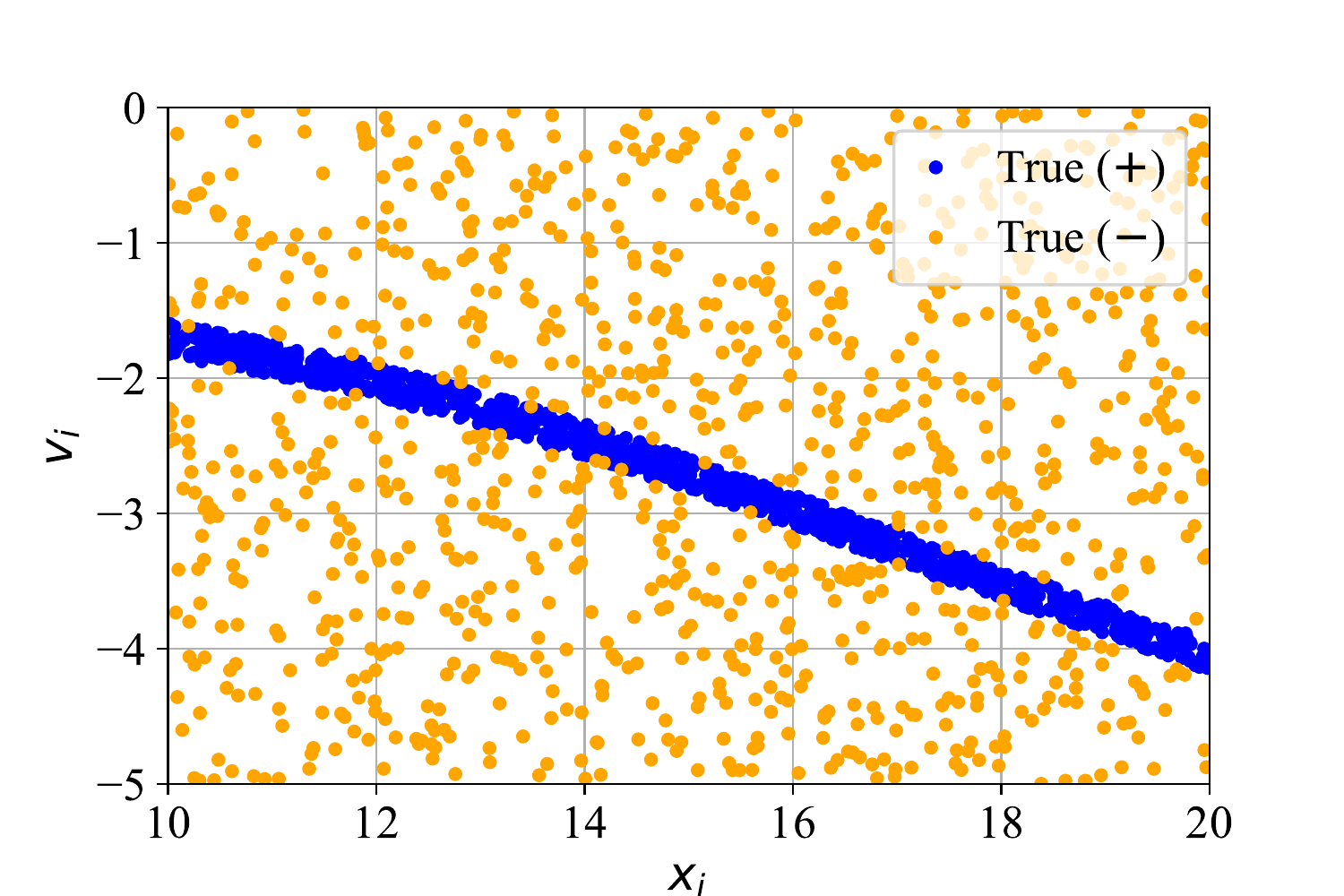}\label{case2inputdataa}}
    \subfigure[$v_{f}$ data for the positive data ($\kappa=0$).]{\includegraphics[width=0.48\textwidth]
    {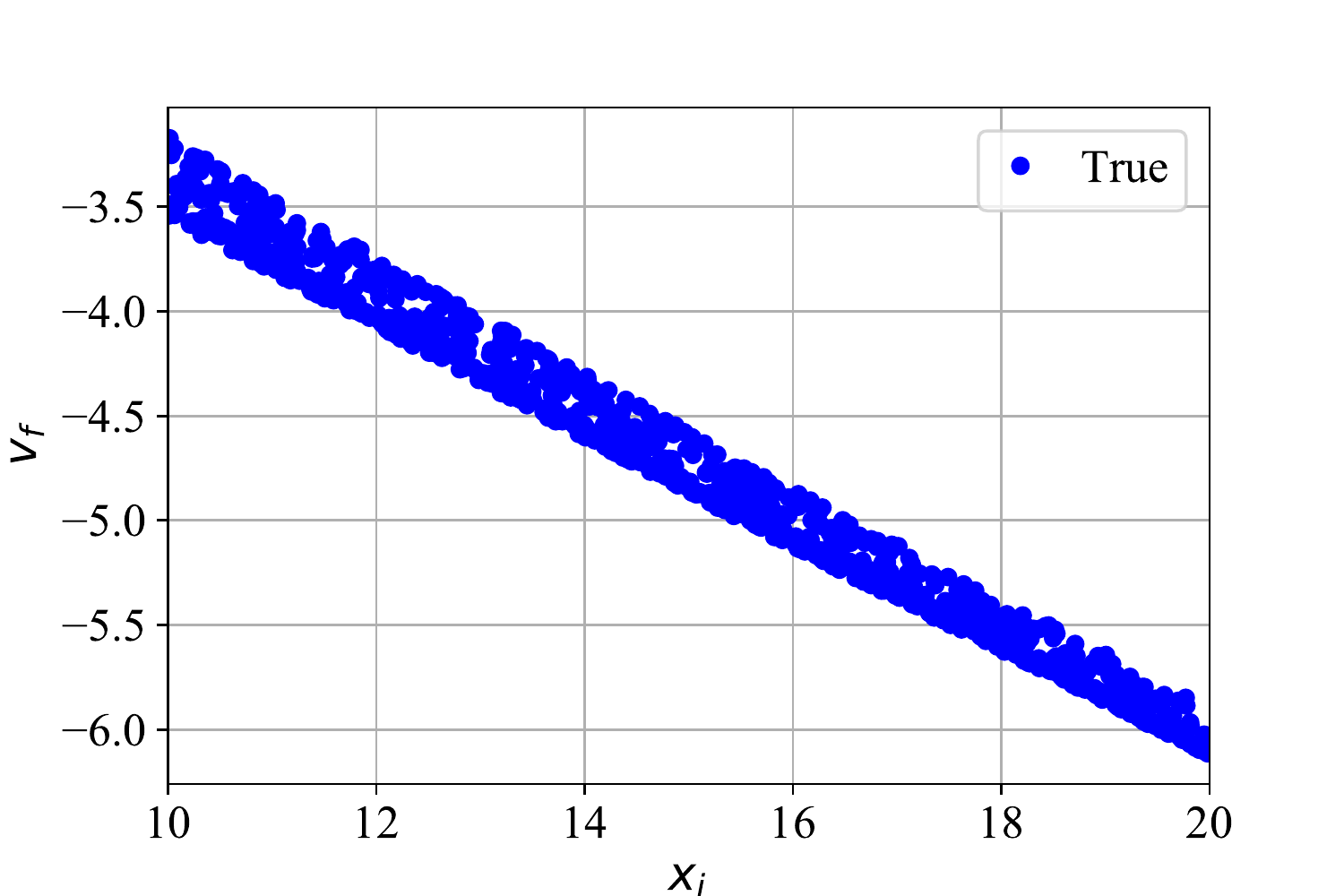}\label{case2inputdatab}}
    \caption{Input data for case 2. }
    \label{fig:F_2_input}
\end{figure}

\paragraph{Examples}
As an example force field, the following hypothetical (complicated) form of $F_{2, \text{True}}$ is assumed and the training data set $\left\{ \left(x_i, v_i, \kappa, v_f \right)\right\}$ is collected.
\begin{equation}
F_{2,\text{True}}(x) = \frac{1}{8000}(x-1)(x-11)^2(x-23)^2-0.7\, ,
\end{equation}
which is shown as the gray line in Fig.~\ref{case2traina}. The experimental input data points $(x_i, v_i)$ are generated uniform-randomly in their preset range and the output data points ($\{\kappa\!=\!0, v_f\}$ for positive data points and $\{\kappa\!=\!1\}$ for negative data points) are collected using an ODE solver with $F_{2, \text{True}}(x)$.
The number of the collected data points for training is 2,000 in total (1,000 for positive, 1,000 for negative). Fig.~\ref{fig:F_2_input} shows the training data points. Fig.~\ref{case2inputdataa} shows the initial kinematic variables $(x_i, v_i)$ where positive ($\kappa\!=\!0$) and negative ($\kappa\!=\!1$) data points are marked differently with blue and orange, respectively. Fig.~\ref{case2inputdatab} shows the distribution of the final velocity $v_f$ of positive data points with respect to the initial position $x_i$.

The learning process and result is put together in Fig.~\ref{fig:F_2_train}. In Fig.~\ref{case2traina} and Fig.~\ref{case2trainb}, the before-learning training data points are put together with the model data points. Fig.~\ref{case2traina} shows the initial kinematic variables $(x_i, v_i)$ of the positive data points ($\kappa\!=\!0$) by blue and the model-positive ($\bar{\kappa}\!\simeq\!0$) by orange where their intersection is marked as green; the green portion shows how much the model is correct on matching $\bar\kappa$ with ${\kappa}$. Note that the intersection of negative and model-negative points ($\kappa\!=\!1$ \& $\bar{\kappa}\!\simeq\!1$) are omitted for clarity.
Fig.~\ref{case2trainb} shows the distribution of the final velocity values from positive data points ($v_f$) by blue and those from model propagation ($\bar{v}_f$) by green, respectively; their discrepancy means the model is not matching $\bar{v}_f$ with $v_f$ correctly. It is clear that the NN model before learning is incorrect on matching either $\bar{\kappa}$ with ${\kappa}$ or $\bar{v}_f$ with $v_f$.

Fig.~\ref{case2trainc} and Fig.~\ref{case2traind} show the ``after-learning'' plots corresponding to Fig.~\ref{case2traina} and Fig.~\ref{case2trainb}, respectively. From the increased portion of green dots in Fig.~\ref{case2trainc} and the accurate matching between blue and green dots in Fig.~\ref{case2traind}, it is clear that the NN model after learning is matching the outputs correctly. Meanwhile, Fig.~\ref{case2traine} shows how $F_2(x)$ is trained over different epochs. It shows how the profile of $F_2$ proceeds from the initial random distribution to the true profile $F_{2,\text{True}}$ by matching $\bar\kappa$ with $\kappa$ and $\bar{v}_f$ with $v_f$ while guided by the regularization error. From these plots, we can see that the NN model matched both sets of output variables correctly as well as accurately discovering $F_{2,\text{True}}$.

\begin{figure}
    \centering
    \includegraphics[width=0.48\textwidth]{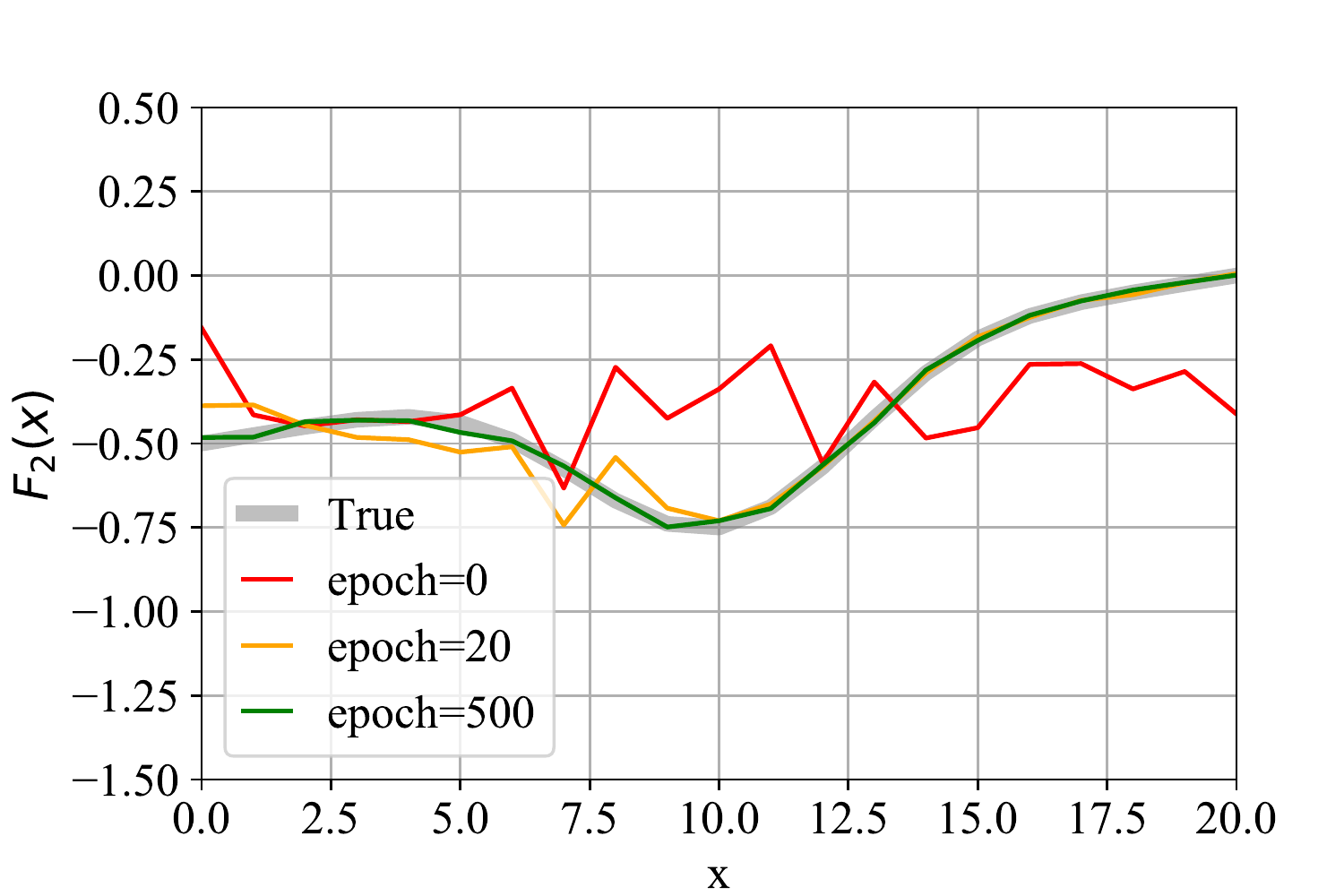}
    \includegraphics[width=0.48\textwidth]{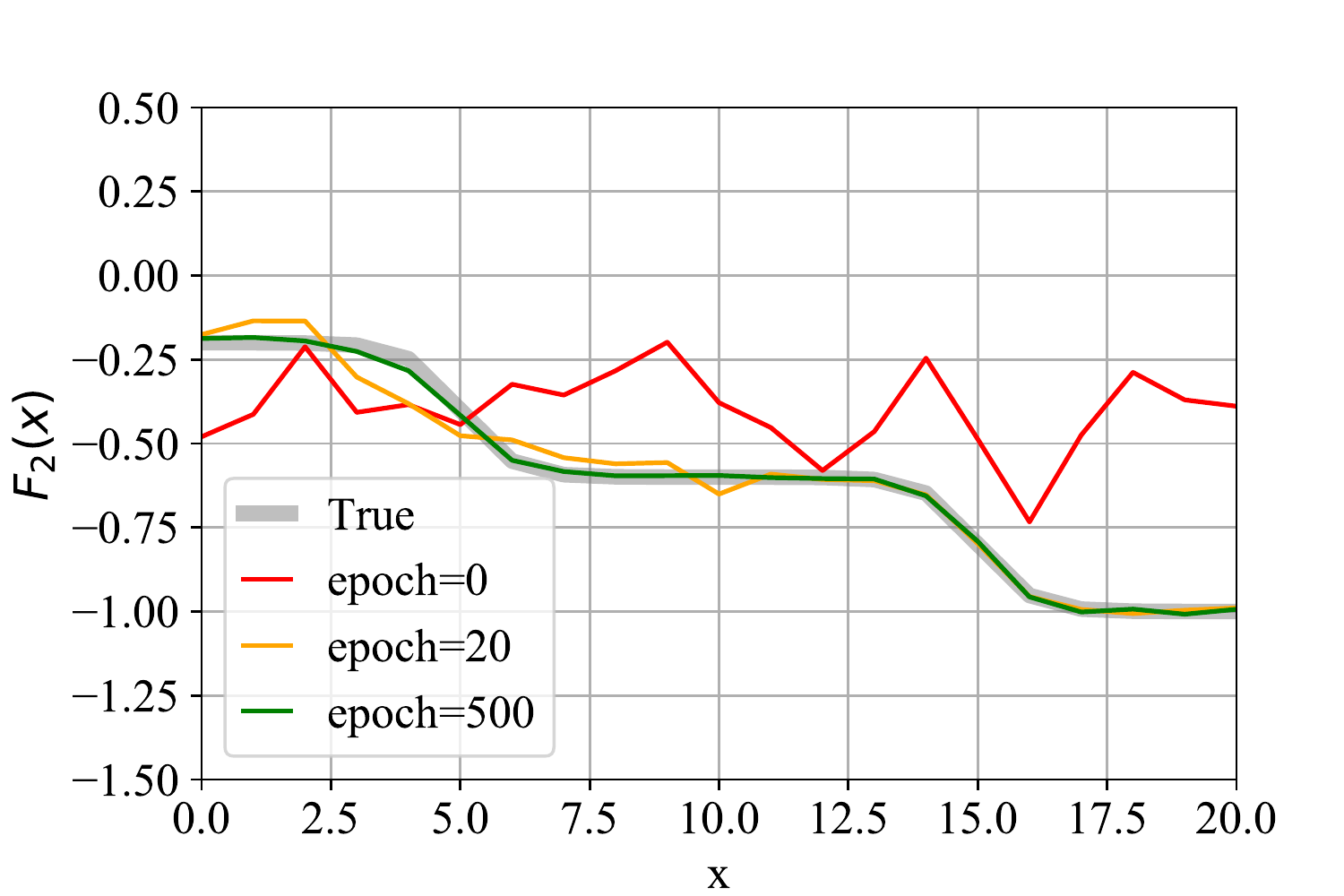}
    \caption{Learning results from different $F_{2,\text{True}}$ profiles.
    }
    \label{fig:my_labelsss}
\end{figure}

To further test the capability of the NN to discover force fields, different-shaped force field profiles are tested with the same scheme. As Fig.~\ref{fig:my_labelsss} shows, the NN discovered the right $F_{2,\text{True}}$ profiles accurately for both cases at $\text{epoch}=500$.\footnote{For the purpose of the test of our method, these force profiles are generated by fitting artificially chosen complicated data. Their functional forms are $\left(\cos\left( \frac{\pi x}{10}\right)-1\right)^{\frac{3}{16}} + (\frac{x}{40}-\frac{1}{2}) $ and
$-\frac{1}{5}\left(\tanh(x-5) + \tanh(x-15) + 3\right)$ respectively.} From this result, we can certify that the NN built with this methodology is capable of learning different shapes of complicated force fields.

\section{Conclusions}

In this paper we analysed classical mechanics problems using the DL.
The main idea of our problem is explained in Fig.~\ref{fig:blackbox}: how to find the unknown force, by a deep learning technique, only from the initial and final data sets.
When the equation of motion (EOM) of a system is given with a set of initial conditions, calculating the propagation of variables numerically is usually not a hard work. On the other hand, retrieving the EOM (or the unknown force in the equations) from a given data set can be a very challenging task, especially with limited types/amount of information (e.g. only initial and final data).

By constructing the NN reflecting the EOM (Fig.~\ref{fig:basicNN}), together with enough input and output data, we successfully obtained the unknown complicated forces. The learning progresses to estimate the unknown forces are shown in Fig.~\ref{case1b},Fig.~\ref{case1others},Fig.~\ref{case2traine}, and Fig.~\ref{fig:my_labelsss}. They show that our DL method successfully discovers the right force profiles without being stuck at local minima, or the multiplicity of mathematically possible solutions.

There are two major advantages of our method.
First, the approach with DL can easily find a complicated answer, which does not allow much intuition to ``correctly guess'' the right form of the answer. 
Second, contrary to usual NN techniques, our approach trains physical quantities such as the unknown force assigned in the NN, which is important for understaning physics.

Our framework can be generalized in a few directions. First, we can consider many particle cases and/or higher dimensional problems. In this case, the number of the kinematic variables increases, which means the width of the NN increases in Fig.~\ref{fig:basicNN}. Second, we can improve our discretization method \eqref{eqn123} by adding higher order corrections or using the neural ODE technique developed in \cite{Hashimoto:2020jug}.  Third, we may apply our method to more complicated problems. For example, we may consider a scattering experiments by unknown forces, which is not a simple power-law force or not even a central force. Last but not least, this work will be pedagogical and heuristic for those who want to apply AdS/DL to the emergence of spacetime as a neural network.


In a broader view, the examples in this paper can enhance the mutual understanding of physics and computational science in the context of both education and research by providing an interesting bridge between them.

\acknowledgments
We would like to thank Koji Hashimoto, Akinori Tanaka, Chang-Woo Ji, Hyun-Gyu Kim, Hyun-Sik Jeong for valuable discussions and comments.
This work was supported by Basic Science Research Program through the National Research Foundation of Korea(NRF) funded by the Ministry of Science, ICT \& Future
Planning (NRF-2017R1A2B4004810) and the GIST Research Institute(GRI) grant funded by GIST in 2020. M. Song and M. S. H. Oh contributed equally as the first author.

\bibliographystyle{JHEP}

\providecommand{\href}[2]{#2}\begingroup\raggedright\endgroup

\end{document}